# Chapter 4

# Wearable Sensors for Individual Grip Force Profiling

## Birgitta Dresp-Langley

### 4.1. Introduction

Biosensors and wearable sensor systems with transmitting capabilities are currently developed and used for the monitoring of health data, exercise activities, and other performance data [1]. Unlike conventional approaches, these devices enable convenient, continuous, and/or unobtrusive monitoring of a user's behavioral signals in real time. Examples include signals relative to body motion, body temperature, blood flow parameters and a variety of biological or biochemical markers and, as will be shown in this chapter here, individual grip force data that directly translate into spatiotemporal grip force profiles for different locations on the fingers and/or palm of the hand. Wearable sensor systems combine innovation in sensor design, electronics, data transmission, power management, and signal processing for statistical analysis, as will be further shown herein. The first section of this chapter will provide an overview of the current state of the art in grip force profiling to highlight important functional aspects to be considered. In the next section, the contribution of wearable sensor technology in the form of sensor glove systems for the real-time monitoring of surgical task skill evolution in novices training in a simulator task will be described on the basis of recent examples. In the discussion, advantages and limitations will be weighed against each other. Finally, although a lot of research is currently devoted to this area, many technological aspects still remain to be optimized, and new

Birgitta Dresp-Langley
Centre National de la Recherche Scientifique (CNRS), UMR 7357 ICube Lab, CNRS and University of Strasbourg, France

methods for data analysis and knowledge representation are urgently needed. These aspects represent an open challenge for the scientific community in the field of wearable sensor technology, as explained in the conclusions of this chapter.

## 4.2. Functional Characteristics of Human Grip Force

Human grip force is controlled at several hierarchical stages, from sensory receptors to the brain and back to the hand, and its functional aspects have been relatively well studied. For example, the relationship between individual finger positions and external grip forces of men and women was investigated [2] in studies where subjects held cylindrical objects from above, using circular precision grips in the 5-finger grip mode. Effects of 4-, 3- and 2-finger grip modes in circular grip mode were also investigated. Individual finger position was nearly constant for all weights and for diameters of 5.0 and 7.5 cm. The mean angular positions for the index, middle, ring and little fingers relative to the thumb were 98 degrees, 145 degrees, 181 degrees, and 236 degrees, respectively. At the 10-cm diameter, the index and middle finger positions increased, while the ring and little finger positions decreased. There were no differences in individual finger position with regard to gender, hand dimension, or hand strength. Total grip force increased with weight, and at diameters greater or lesser than 7.5 cm. Total grip force also increased as the number of fingers used for grasping decreased. Although the contribution of the individual fingers to the total grip force changes with object weight and diameter, the thumb contribution always exceeded 38 % followed by the ring finger and the little finger (pinky), which contributed approximately 18-23 %, for all weights and diameters. The contribution of the index finger was always smallest, and there was no gender difference for any of the grip force variables. Effects of hand dimension and hand strength on the individual finger grip forces were subtle and minor [2].

The contributions and co-ordination of external finger grip forces during a lifting task with a precision grip using multiple fingers were also investigated [3]. Ten subjects lifted a force transducer-equipped grip apparatus, and grip force from each of the five fingers was continuously measured with different object weights and surface structures. Effects of five-, four-, and three-finger grip modes were also examined [3]. It was found that variation of object weight or surface friction resulted in a change of the total grip force magnitude. The largest change in finger force was recorded for the index finger, followed by the middle, ring, and little fingers. Percentage contributions of static grip force to total grip force for the index, middle, ring, and little fingers was 42.0 %, 27.4 %, 17.6 % and 12.9 %, respectively, and values were roughly constant across object weights and surface friction conditions. These results suggest that all individual finger force adjustments for lighter loads (<800 g) are controlled by using a single common scaling value [3]. Higher surface friction provided faster lifting initiation and required lesser grip force, indicating the beneficial effects of a non-slippery surface. Nearly 40 % force reduction was obtained with the non-slippery surface. Variation in grip mode changes the total grip force: the fewer the number of fingers used, the greater is the total grip force [3]. The static grip force for the index, middle, and ring fingers in the four-finger grip mode was 42.7 %, 32.5 %, and 24.7 %, respectively. Static grip force for the index and middle fingers in the three-finger grip mode was 43.0 % and 56.9 %, respectively, suggesting that the grip mode, i.e. whether all five or only three fingers are used, influences the force contributions of the middle and ring fingers, but not that of the index finger [3].

Other studies have shown a phenomenon of motor redundancy [4] in human prehensile behavior [4]. The partly redundant design of the hand allows performing a variety of tasks in a reliable and flexible way following the principle of abundance, as shown in robotics with respect to the control of artificial grippers, for example. Multi-digit synergies appear to operate at two levels of hierarchy to control prehensile action [4]. Forces and moments produced by the thumb and the "virtual finger" (an imagined finger with a mechanical action equal to the combined mechanical action of all other four fingers of the hand) co-vary at a higher level only to stabilize the grip action in respect to the orientation of the hand-held object. Analysis of grip force adjustments during motion of hand-held objects suggests that the central nervous system adjusts to gravitational and inertial loads differently, at an even higher level of control. Object manipulation by efficient control of finger and grip force is therefore not only a motor skill but also a cognitive skill [5] exploited in surgery, craft making, and musical performance. Sequences of relatively straightforward cause-effect links directly related to mechanical constraints lead on to a non-trivial co-variation between low-level and high-level control variables [3-5], as in playing a musical instrument, which requires independent control of the magnitude and rate of force production, which typically vary in relation to loudness and tempo [5].

The expert performance of a highly skilled pianist, for example, is characterized by a rapid reduction of finger forces, allowing for considerably fast performance of repetitive piano keystrokes [5]. Skilled grasping behavior (multi-finger grasping) has three essential components: 1) manipulation force, or resultant force and moment of force exerted on the object and the digits' contribution to force production, 2) internal forces, which are defined as forces that cancel each other out to maintain object stability, to ensure slip prevention, tilt prevention, and robustness against perturbation, and 3) motor control (or grasp control), which involves prehensile synergies, chain effects, inter-finger connection [2-4] and the high-level brain command of simultaneous digit adjustments to several, mutually reinforcing or conflicting demands. Prehensile synergies are reflected by characteristics of digit action and their co-variation patterns during task execution or during static holding of an object while the external torque force changes slowly and smoothly.

Conditions of torque forces changing slowly, requiring smooth adjustments of grip efforts from either non-zero pronation to zero, or from non-zero supination to zero were investigated [7]. With the handle kept vertical at all times, indices of variance and co-variation of forces and moments of force produced by individual digits to stabilize performance were measured in terms of total normal force, total tangential force, and total moment of force. Measurements were computed at two levels of an assumed control hierarchy: 1) an upper level, where the task is shared between the thumb and the "virtual finger", i.e. an imagined digit with mechanical action equal to that of all other four fingers and 2) a lower level, where the action of the "virtual finger" is shared by the true four fingers. When total moment of force is expressed in terms of the sum of the moments of force produced by the thumb and the "virtual finger", or the sum of the moments of force produced by normal forces and tangential forces, it was found that adjustments in total moment of force were produced primarily by changes in the moment produced by the "virtual finger" and by changes in the moment produced by the normal forces. The normal force of the thumb at the final state was the same across conditions, and solely determined by changes in the external torque force [7]. The co-contraction index reflecting the moment of force production by true four fingers acting against the total moment produced by the "virtual finger" was higher in the "from non-zero supination to zero" condition. Variance indices dropped with the decrease in external torque force while co-variation indices remained unchanged over the task time [7].

These results suggest a trade-off between the two levels of hierarchy assumed, with larger indices at the higher level corresponding to smaller indices at the lower level as characteristic features of prehensile tasks. Functional properties of digit action and interaction do not only depend on the magnitudes of external constraints (i.e. external torque forces), but also on temporal changes in such constraints and their history. Static grasping/holding of a horizontally oriented object was also explored to address issues relative to the sharing patterns of the total moment of force across the digits, the presence of mechanically unnecessary digit forces, and the trade-off between multi-digit synergies at the two levels of assumed control hierarchy [8]. Measurement conditions consisted of holding statically a horizontally oriented handle instrumented with six-component force/torque sensors with different loads and torques acting about the longer axis of the handle. The thumb acted from above, while the other four fingers supported the weight of the object. When the external torque force is zero, the thumb produces a mechanically unnecessary force which does not depend on the external load magnitude [8]. When the external torque force is non-zero, the tangential forces produced over 80 % of the total force. The normal forces by the middle and ring fingers produce consistent moments against the external torque force, while the normal forces of the index and little fingers do not [8]. Overall, results have shown that task mechanics are only one of the many factors that determine the grip forces produced by individual digits. Sensory receptor processing may lead to mechanically unnecessary forces [7, 8], and there seems to be no single rule that describes the sharing of normal and tangential forces across tasks. Fingers such as the ring finger are traditionally viewed as "less accurate" [1-3], yet they may perform more consistently in specific tasks [8].

The trade-offs between variables produced at the two hierarchical control levels assumed suggest that a degree of "functional redundancy" at the higher control level represents an important characteristic of grip force control. Other studies measured grip strength using several methods serially excluding one or two phalanges [9] to shed further light on clinical and/or functional outcomes relating to the contribution of each finger to overall grip force. Two hundred healthy young men were included in this survey, and demographic variables as well as anthropometric parameters of forearms and hands were recorded. Grip strength was measured using all fingers, all fingers except the thumb, all fingers except the index finger, all fingers except the middle finger, all fingers except the ring and little fingers, and all fingers except the little finger. The contribution of each finger to total grip strength was estimated [9]. Grip strength using all five fingers was, as would be expected, greatest, closely followed by grip strength without the thumb. Grip strengths without the middle, the ring and the little finger were the smallest. Contributions of the index, middle, ring and little fingers to the grip strength were 17 %, 22 %, 31 %, and 29 %, respectively. The middle finger was the most important contributor to grip strength, followed by the combination of ring and little fingers. Positive correlations between each grip method and anthropometric parameters such as hand size were found [9]. Grip force and its distribution across different fingers of the hand are important parameters for evaluating functional aspects of grip forces during task execution. Identifying differences between non-dominant and dominant hands in bimanual grip tasks as a function of anthropometric parameters is another potentially important source of variation.

Some authors [10] compared grip force and load distributions of dominant and non-dominant hands in right-handed healthy subjects, assigned to either a small or a large cylindrical object with respect to their hand size. Maximum and mean grip forces well as the contribution (in percent) of each digit, *thenar,* and *hypothenar* in relation to the total load applied were measured and compared across the dominant and non-dominant hands. The contribution to mean grip strength differed significantly between the thumb and the ring finger, the thumb and the little finger, and between thumb and measurements taken from the palm of the hand. The dominant hand showed a smaller force contribution of the thumb and ring finger, and a greater contribution of the palm of the hand in comparison with the non-dominant hand [10].The contribution of the small fingers to maximum grip force was equal between the dominant and non-dominant hands; no differences were found between the index finger, middle finger, *thenar*, and *hypothenar* when analyzing their cumulated % force contribution to both mean and maximum force. In right-handed subjects, the thumb and the ring finger are functionally the most important contributors to grip force.

## 4.3. Wearable Grip Force Sensors for Individual Grip Force Profiling in Real Time during Bimanual Task Execution

The example of surgical task training and, in particular, robot assisted minimally invasive surgical training, is evoked here to bring forward clinical, ergonomic, and general functional advantages of individual grip force profiling using wearable (gloves or glove-like assemblies) sensor systems for the monitoring of task parameters relating to manual skill evolution in real time. In studies on robotic surgery platforms [11, 12], for example, the bi-manual performance skill learning curves of experienced urological robotic surgeons, surgeons with experience as robotic platform tableside assistants, urological surgeons with laparoscopic experience, urological surgeons without laparoscopic experience, and complete novices, either aged 25 and younger, or 40 and older were compared. The results showed that all experienced robotic surgeons reached expert performance level (>90 %, as defined previously in the literature) within the first three trial repeats, and consistently maintained a high level of performance. All other groups performed significantly worse. Platform tableside assistants, laparoscopy experienced surgeons, and younger novices showed better performance in all exercises than surgeons without laparoscopic experience and older novices. In summary, performance in robotic surgery measured by performance scores in virtual simulator modules is significantly dependent on age, and on prior experience with robotic and laparoscopic surgery [12]. Minimally invasive robotic surgery has many advantages over traditional surgical procedures, but the loss of force feedback yields potential for stronger grip forces during task execution, which can result in excessive tissue damage [13].

Grip force monitoring is therefore a highly useful means of tracking the evolution of the surgeon's individual force profile during task execution [14]. While current multi-modal feed-back systems may represent a slight advantage over the not very effective traditional single modality feedback solutions by achieving average grip forces closer to those normally possible with the human hand [13], the monitoring of individual grip forces of the surgeon (or trainee) during task execution by wearable multisensory systems is by far the superior solution.

Real-time grip force sensing by wearable systems can directly help prevent incidents, because it includes the possibility of sending a signal (sound or light) to the surgeon whenever his/her grip force exceeds a critical limit before the damage is done. Proficiency, or expertise, in the control of a robotic master/slave system designed for minimally invasive surgery is reflected by a lesser grip force during task execution as well as by a shorter task execution times [14, 15]. Benchmark measures permitting to establish objective criteria for expertise in using such surgical systems, which have a limited degrees of freedom, need to be found to ensure effective training of future surgeons. As shown in the introduction here above, the state of the art in experimental studies on grip force strength and control for lifting and manipulating objects strategically has provided limited insight into the contributions of each finger to overall grip strength and fine grip force control, with a general conclusion in terms of complex prehensile synergies governed by interactions between at least two hierarchical processes of sensory (low-level) and central (high-level) information processing and control. However, what generally has emerged from that research is that 1) the middle finger is the most important contributor to the gross total grip force and, therefore, most important for getting a good grip of heavy objects to lift or carry, while 2) the ring finger and the small (pinky) finger appear most important for the fine control of subtle grip force modulations [16, 17], which is important in surgical tasks, and even more critically important when manipulating the handles of a surgical robot with limited degrees of freedom for hand and finger movements. Also, grip force may be stronger in the dominant hand compared with the non-dominant hand. In two recent studies [14, 15] the grip force profiles corresponding to measurements collected from specific sensor positions on anatomically relevant parts of the finger and hand regions of the dominant and non-dominant hands of an expert in controlling a robotic surgery system were compared to those of a beginner, who manipulated the device for the first time.

A wireless sensor glove hardware-software system, described in detail in [14, 15], was specially designed for these studies. Before the sensor glove system was employed to study expert and novice grip force profiles during manipulation of the robotic system, individual grip force data from the right and left hands of several young individuals were recorded in preliminary test sessions using weighted handles, which had to be lifted up and down to the sound of different types of music [18]. These systematic tests produced relevant data relative to the characteristics of individual grip force profiles in time as a function of the sensor location on the fingers and palms of either of the two hands (Fig. 4.1, left), the hand considered (dominant *versus* non-dominant), and the type of music played during two-handed manipulation (soft music *versus* hard rock music) of the weighted handles. Results from these preliminary tests are described in the next section here below to illustrate how statistical analyses of the individual grip force profiles shed light on specific aspects of individual task performance.

The wireless gloves (Fig. 4.1, right) that produced the data shown here below were specifically designed for individual grip force profiling [14, 15, 18]. They contain 12 small force sensitive resistors FSR, in contact with specific locations on the inner surface of the hand as given in Fig. 4.2.

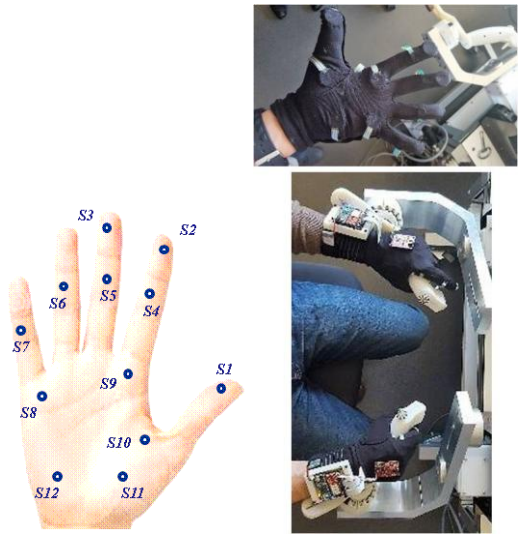

**Fig. 4.1.** Signals relative to grip force were sampled from 12 anatomically relevant force sensitive sensor (FSR) locations on the fingers and in the palm (left) of both hands. The FSRS were sewn into a soft glove (top right) wireless wearable sensor system design [14, 15, 18] for grip force monitoring in bimanually executed tasks such as robotic surgery (bottom right).

Two layers of cloth were used and the FSRs were inserted between the layers. The FSRs did not interact, neither directly with the skin of the subject, nor with the master handles, which provided a comfortable feel when manipulating the system. FSRs were sewn into the glove with a needle and thread. Each FSR was sewn to the cloth around the conducting surfaces (active areas). The electrical connections of the sensors were individually routed to the dorsal side of the hand and brought to a soft ribbon cable, connected to a small and very light electrical casing that was strapped onto the upper part of the forearm and equipped with an Arduino microcontroller. Eight of the FSR positioned in the palm of the hand and on the finger tips had a 10 mm diameter, while the remaining four located on the middle phalanxes on the fingers had a 5 mm diameter. Each FSR was soldered to 10 KΩ pull-down resistors to create a voltage divider. The voltage (V) read by the analog input of the Arduino is given by (4.1)

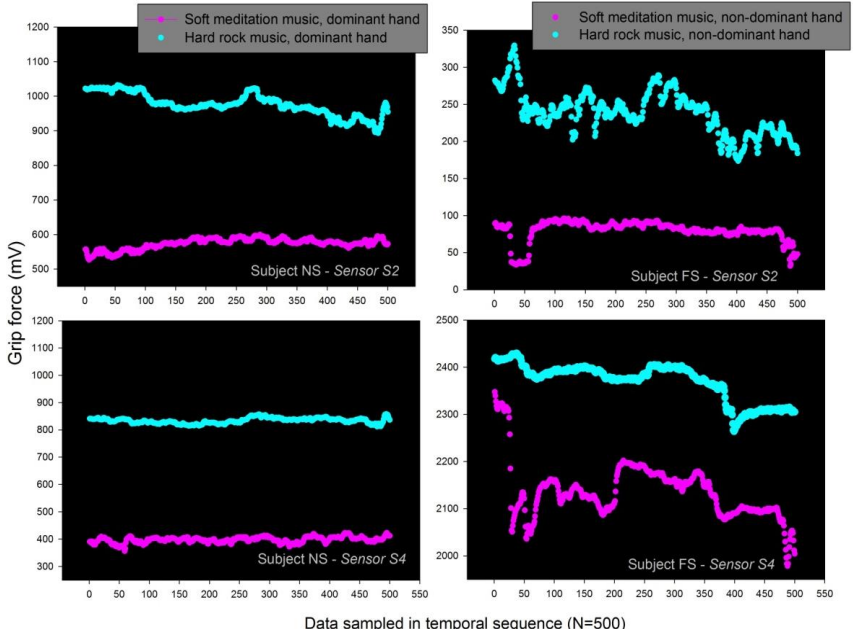

**Fig. 4.2.** Effects of sound (music) in sensor locations S2 and S4 recorded in the temporal task sequences of two individuals (NS, left; FS, right).

$$Voutput = RPDV3.3/(RPD+RFSR), \qquad (4.1)$$

where RPD is the pull-down resistance, RFSR the FSR resistance, and 3.3 the V supply voltage. FSR resistances can vary from 250 Ω when subject to 20 Newton (N) to more than 10 MΩ when no force is applied at all. Voltages generated in the experiments from which data were drawn [14, 15, 18] varied monotonically between 0 and 3.22 V as a function of grip force applied, which is assumed uniform on the sensor surface. In the experiments, grip forces did not exceed 10 Newton, and voltages varied within the range of [0; 1500] mV. The relation between force and voltage is almost linear within this range. All sensors provided similar calibration curves and comparisons could be made directly between voltages at the millivolt scale. Regulated 3.3 V was provided to the sensors through the Arduino. Power was provided by a 4.2 V Li-Po battery enabling wireless use of the glove system. The battery voltage level is directly controlled by the Arduino, and displayed continuously on the screen of the user interface. The glove system was connected to a computer for data storage via Bluetooth enabled wireless communication running 115,200 bits-per-second (bps). The software of the glove system has two parts: One running on the gloves, and one running on the computer algorithm for data collection. Each of the two gloves is sending data to the computer separately, and the software reads the input values, and stores them on the computer according to header values indicating their origin. The software running on the Arduino was designed to acquire analog voltages provided by the FSR every 20 milliseconds (50 Hz). Input voltages are merged with their time stamps and sensor identification. The data package is sent to the computer via Bluetooth, which is decoded by the computer software. The voltages are saved in a text file for each sensor, with their time stamps and identifications. The computer software monitors the voltage values received from the gloves via a user interface displaying battery levels. Detailed descriptions and images of the wireless sensor glove in action on a robotic surgery system, system specifications, and the general design chart of the hard-to-software wireless wearable sensor system have been published in [14], accessible online at https://www.mdpi.com/1424-8220/19/20/4575/htm (Licence CC BY 4.0).

## 4.3. Data Analysis and Visualization

The individual grip force profiling and the group profiles analyzed and visualized here correspond to unpublished data from the preliminary testing phase of the wireless sensor gloves [18] recorded from young volunteers while moving two weighted handles up and down to the sound of different pieces of music. The handles subjects were commercially available cylindrical weights of identical shape and size weighing one kilogram each. Different pieces of music were selected for the different sound exposure conditions. Exposure duration was 10 seconds for each of them. One piece consisted of extremely soft tones designed for meditation, another of highly aggressive hard metal rock performed by the group *Rammstein* (*"Zerstören"* from their album *"Rosenrot"*). Sound intensities were maintained the same for the different pieces of music, on the computer and on the two loudspeakers. Nine healthy men and two healthy women, aged between 20 and 30, all of them right-handed, participated in this study. Handedness was confirmed individually using the Edinburgh inventory for handedness. The subjects were all volunteers and naive to the purpose of the experiment. The study trials were conducted in full conformity with the Helsinki Declaration relative to scientific experiments on human individuals and approved by the ethics board of the lead investigator's host institution (CNRS). All participants were young volunteers and provided written informed consent. Their identity is not revealed. Hand grip forces were recorded from the twelve sensor *loci* on the dominant and non-dominant hands of eleven subjects in different experimental conditions [18]. All subjects were tested in all the conditions of exposure to different types of music. The order of these different conditions was carefully counterbalanced between subjects. During the tests, subjects were standing upright facing a table on which the two handles they had to grip were placed in alignment with the forearm motor axis. Subjects were instructed to grab the handles with their two hands and to start moving them up and down as soon as the music started. The duration of the music and the handgrip force recordings was ten seconds per subject and experimental condition. Raw data (voltages) from each sensor were recorded every 20 milliseconds in a temporal sequence, for each subject and experimental condition, committed to *Excel* files with labeled columns. Grouped data were imported into *Matlab 7.14* for transformation of the voltage output (*Voutput*) data into Newton (*N*) by (4.2)

$$N = Voutput/(RPDV3.3\text{-}Voutput), \qquad (4.2)$$

where RPD is the pull-down resistance, RFSR the FSR resistance, and 3.3 the V supply voltage. The individual force profiles here below show grip force data in millivolt (mV), the grouped force profiles show population grip force data in Newton. Either parameter is valid, given that for the range of variations recorded in the different experiment, the relation between force and voltage is almost linear.

The analyses here below concern individual grip force profiles from two sensor locations (S2 and S4) of two individuals from a subject pool of eleven. Grip forces recorded in task time were plotted as a function of two different music conditions, and the hand they were recorded from. In the grip force profiles of the two individuals, we see that S2 and S4 data produces a clear effect of music (sound), with stronger grip forces for hard rock music. These profiles (Fig. 4.2) also show that the two individuals apply different total forces at the two sensor locations.

The sensors S2 and S4, among others, also showed differences in forces applied by the dominant and the non-dominant of some individuals. Individual grip force profiles of the dominant and the non-dominant hands recorded from these two sensor locations are shown here below for subject FS in the hard rock music condition. The individual grip force profiles reveal systematically stronger grip forces applied by the dominant hand in sensor locations S2 (Fig. 4.3, top) and S4 (Fig. 4.3, bottom), and systematically and considerably stronger total grip forces in sensor S4, when comparing the range of variations shown (top graph *versus* bottom graph).

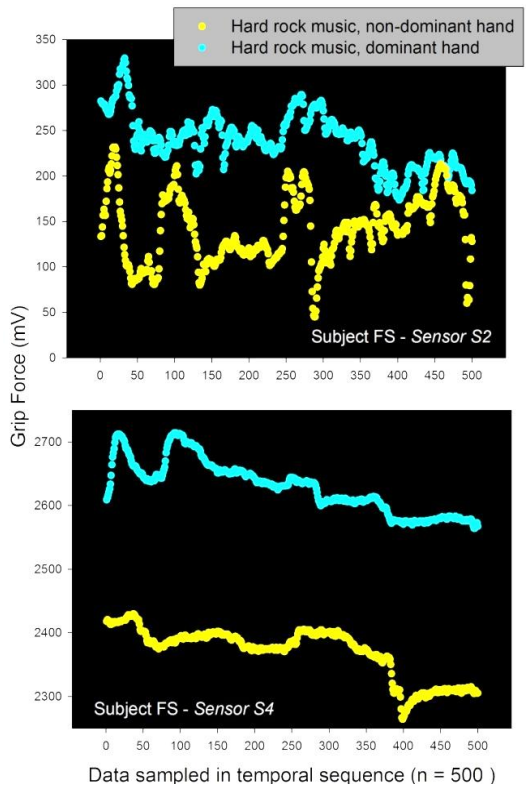

**Fig. 4.3.** Effects of hand on grip forces recorded from sensor locations S2 and S4 recorded in the temporal task sequence of individual FS in the hard rock music condition.

Average grip force data of a whole population of individuals in a given condition, or factorial combination thereof, may also be computed. This allows to assess how individual grip force profiles compare with the average profile of a whole population under the same conditions. Examples of average population grip force data for the dominant and the non-dominant hands from different sensor locations and across music conditions are shown as examples (Fig. 4.4) here below.

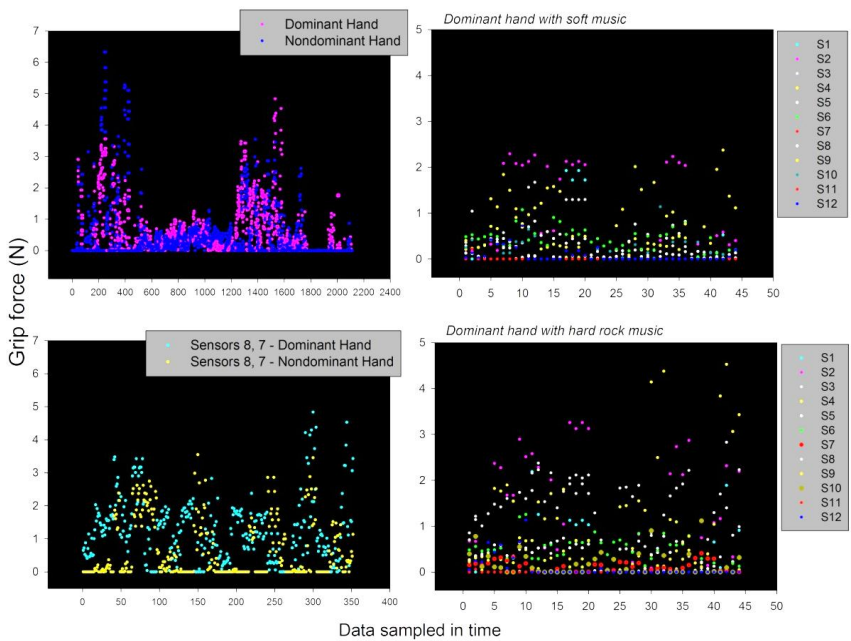

**Fig. 4.4.** Effects of hand, sensor location, and sound (music) on average population grip forces computed on data sampled in time from the whole study population.

The examples here above show that combining wireless wearable multisensory technology with an appropriate experimental design produces meaningful individual and general (population) grip force profiles that can be compared for the selection of conditions for task performance benchmarking recorded using. The pretesting data (not all are shown here) from these preliminary experiments here have enabled the design of the study on the robotic surgery system [14] referred to here above (with hypertext link). That study produced highly significant differences in expert and novice grip force profiles, as revealed by a series of robust statistical analysis (ANOVA) across surgical task sessions in time and sensor locations [14]. The characteristic individual grip force profile of an expert surgeon during bimanual manipulation of the robotic surgery device (Fig. 4.1) across repeated task sessions for a four-step pick-and-place task [14] displays higher values for subtle grip control by the small finger, and lesser values for gross grip force by the middle finger when compared with the grip force profiles of the novice (Fig. 4.5). Grip force profiling has thereby permitted to benchmark fine grip force control by the small finger and minimal gross force deployment by the middle finger, in combination with shorter task execution times [14], as a typical functional characteristic of surgical expertise on the specific robotic system.

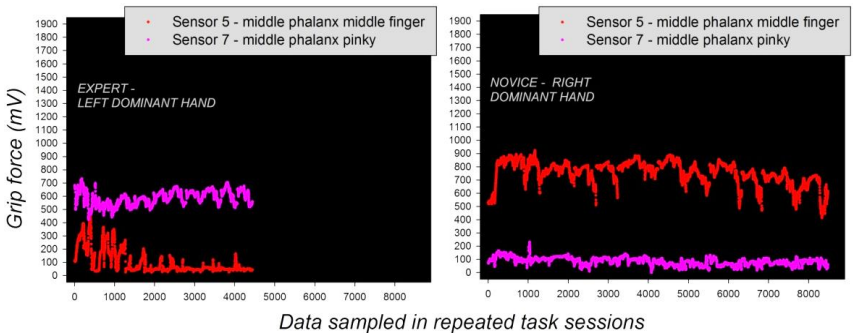

**Fig. 4.5.** Individual grip force profiles of the dominant hand distinguishing an expert's (left) from a novice's (right) performance during bimanual use of a robotic surgery system in four-step-pick-and-drop task [14]. The shorter task execution times per session of the expert (left) are reflected here by the smaller number of data sampled in task time across sessions (x-axis of the graph on left).

These insights, combined with the effects of sound on individual grip force profiles shown from the pretest study here suggest that it should be possible to effectively use individual grip force profiling with sound feed-back in robotic surgery for modulating excessive grip forces during interventions well before they may cause potentially dangerous tissue damage.

### 4.4. Conclusions

Prehensile synergies of the human hand are under the command of multiple levels of sensorial integration, cognitive control, and interactions between the two, as summarized here in the introduction. Using wireless wearable sensor technology, possibly in combination with sensory feed-back systems [19-23], for the effective monitoring of manual and bimanual tasks where grip force matters critically represents a promising way towards performance quality benchmarking in training, and in risk prevention, especially for critical tasks such as robot assisted surgery [14, 15]. The human hand has evolved [24] as a function of active constraints [25-29], and in harmony with other sensory systems such as the auditory system [18, 23, 30] Grip force profiles are a direct reflection of the complex low-level, cognitive, and behavioral synergies this evolution has produced.